\documentclass[twocolumn,showpacs,preprintnumbers,superscriptaddress,aps,prl]{revtex4-1}

\usepackage{graphicx,color}

\usepackage{dcolumn}

\usepackage{bm}

\begin{document}

\preprint{}

\title{Magnetic Dispersion and Anisotropy in Multiferroic BiFeO$_3$}

\author{M. Matsuda}

\affiliation{Quantum Condensed Matter Division, Oak Ridge National Laboratory, Oak Ridge, Tennessee 37831, USA}

\author{R. S. Fishman}

\affiliation{Materials Science and Technology Division, Oak Ridge National Laboratory, Oak Ridge, Tennessee 37831, USA}

\author{T. Hong}

\affiliation{Quantum Condensed Matter Division, Oak Ridge National Laboratory, Oak Ridge, Tennessee 37831, USA}

\author{C. H. Lee}

\affiliation{National Institute of Advanced Industrial Science and Technology (AIST), Tsukuba, Ibaraki 305-8562, Japan}

\author{T. Ushiyama}

\affiliation{National Institute of Advanced Industrial Science and Technology (AIST), Tsukuba, Ibaraki 305-8562, Japan}

\author{Y. Yanagisawa}

\affiliation{National Institute of Advanced Industrial Science and Technology (AIST), Tsukuba, Ibaraki 305-8562, Japan}

\author{Y. Tomioka}

\affiliation{National Institute of Advanced Industrial Science and Technology (AIST), Tsukuba, Ibaraki 305-8562, Japan}

\author{T. Ito}

\affiliation{National Institute of Advanced Industrial Science and Technology (AIST), Tsukuba, Ibaraki 305-8562, Japan}

\date{\today}

\begin{abstract}

We have determined the full magnetic dispersion relations of multiferroic BiFeO$_3$. In particular, two excitation gaps originating from magnetic anisotropies have been clearly observed.
The direct observation of the gaps enables us to accurately determine the Dzyaloshinskii-Moriya (DM) interaction and the single ion anisotropy. The DM interaction supports a sizable magneto-electric coupling in this compound.

\end{abstract}


\maketitle

Multiferroic materials, in which spontaneous ferroelectric polarization and magnetic order coexist, have been investigated intensively not only due to their potential industrial applications but also due to purely scientific interest about magneto-electric coupling in strongly correlated electron systems.  For many geometrically frustrated magnets, ferroelectricity is mediated by the magneto-electric coupling. Several mechanisms have been proposed to explain the spin-driven ferroelectricity. \cite{arima11}

\begin{figure}
\includegraphics[width=7.0cm]{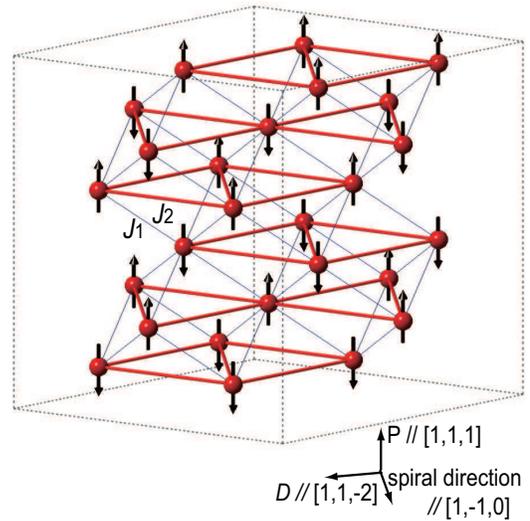}
\caption{(Color online) Magnetic structure of the Fe$^{3+}$ moments in BiFeO$_3$. The basic structure of collinear G-type, in which nearest-neighbor spins align antiferromagnetically, is shown. The dotted lines correspond to the 2$\times$2$\times$1 unit cells of the rhombohedral structure ($R3c$). The pseudo cubic unit cell that is used in this paper is shown by the thin solid lines. The spin structure is spiral along the $[1,-1,0]$ direction with a long period of $\sim$640 \AA. The nearest-neighbor ($J_1$) and next-nearest-neighbor interactions ($J_2$) as well as the directions of the DM vector ($D$) and the spontaneous electric polarization ($P$) are also shown.}
\label{structure}
\end{figure}

BiFeO$_3$ has a rhombohedral structure ($R3c$) below $\sim$1100 K, where ferroelectricity appears.~\cite{teague70} (This paper employs pseudo-cubic notation with $a$$\sim$3.96 \AA\ and $\alpha$$\sim$89.4$^\circ$.)
The ferroelectricity is considered to primarily originate from displacements of the Bi$^{3+}$ ions due to the lone 6$s^2$ pair.
Cycloidal magnetic order with the propagation vectors $\tau_1=(\delta,-\delta,0)$, $\tau_2=(\delta,0,-\delta)$, and $\tau_3=(0,-\delta,\delta)$ develops below $T\rm_N\sim$ 640 K, as shown in Fig. \ref{structure}. \cite{sosnowska82,lebeugle08,slee08} The magnetic structure persists down to low temperatures, although the imcommensurability $\delta$ changes from $\sim$0.0045 at 5 K to $\sim$0.0037 at 600 K. \cite{ramazanoglu11,JHA10,sosnowska11}
Because $T\rm_N$ is much higher than room temperature and because of the large spontaneous electronic polarization ($P\sim$100 $\mu$C/cm$^2$), \cite{shvartsman07,lebeugle07} this material has attracted many researchers and has been studied extensively.~\cite{catalan09}

Although $T\rm_N$ is much lower than the ferroelectric Curie temperature in BiFeO$_3$, several measurements show sizeable magneto-electric coupling.  For example, the magnetic domain distribution can be controlled by applying an electric field. \cite{lebeugle08,slee08,slee08_2} An abrupt decrease (up to $\sim$40 nC/cm$^2$) in electric polarization was also observed in a magnetic field of about 20 T, \cite{kadomtseva04,park11,tokunaga10} where a transition from the incommensurate cycloidal structure to an almost commensurate structure with a weak ferromagnetic component is suggested. \cite{ohoyama11} These results suggest that the additional polarization below $T\rm_N$ is driven by the magneto-electric effect. However, it is important to clarify this mechanism from a microscopic point of view. Very recently, some inelastic neutron scattering studies have measured the spin-wave excitations using powder or single crystal sample. \cite{sosnowska09,delaire12,jeong12}
Furthermore, the detailed spin Hamiltonian including the magnetic anisotropy was discussed to explain magneto-electric coupling in this compound by Sosnowska {\it et al.} \cite{sosnowska95} and Jeong {\it et al.} \cite{jeong12}.
However, a detailed analysis based on the direct observation of the magnetic anisotropy, which is necessary to discuss the magneto-electric coupling, has not yet been performed.

We performed inelastic neutron scattering experiments on a single crystal of BiFeO$_3$. We have determined the full magnetic dispersion relations of the spin-wave excitations in this compound. In particular, low-energy gapped excitations have been detected for the first time by high energy resolution experiments. The direct observation of the excitation gaps makes it possible to determine accurately the magnetic anisotropies due to Dzyaloshinskii-Moriya (DM) interaction and single ion anisotropy. Our detailed analysis has revealed that the coupling constants are $J_1$=6.48 meV, $J_2$=0.29 meV, $D$=0.1623 meV, and $K$=0.0068 meV, where $J_1$, $J_1$, $D$, and $K$ are nearest-neighbor, next-nearest-neighbor, DM interactions, and single ion anisotropy, respectively, which reproduce the cycloidal spin structure.
We also measured the temperature dependence of the dispersion relations below room temperature. Although no drastic change was observed, the spin-wave are slightly softened above 200 K.

\begin{figure}
\includegraphics[width=8.6cm]{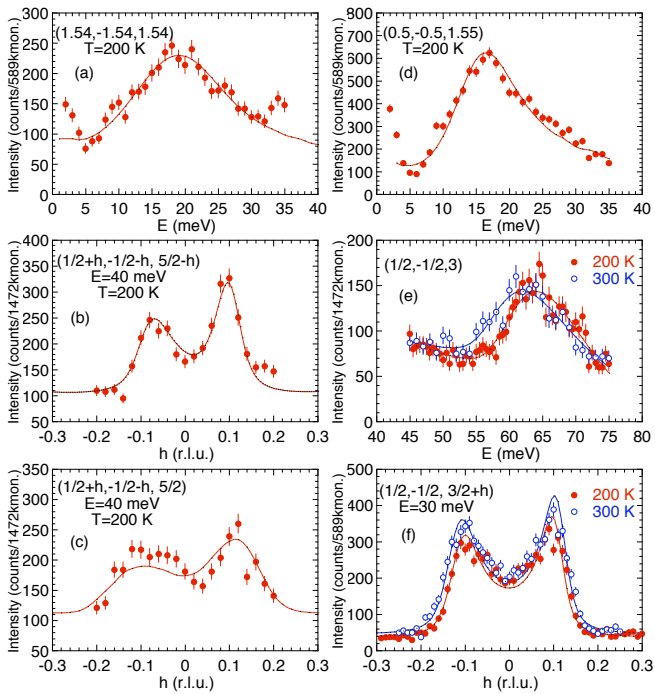}
\caption{(Color online) Typical constant-$Q$ and constant-energy scans along $[h,\bar{h},h]$, $[h,\bar{h},\bar{h}]$, $[h,\bar{h},0]$, and $[0,0,h]$ measured on HB-1 using thermal neutrons at $T$=200 and 300 K in BiFeO$_3$. The solid curves are the results of fits of a convolution of the resolution function with Lorentzians.}
\label{spectra}
\end{figure}
A single crystal of BiFeO$_3$ was grown using the traveling solvent floating zone (TSFZ) method by laser heating, as described in Ref. \onlinecite{ito11}. The dimensions of the single crystal is $\sim$5$\phi\times$40 mm$^3$.  The effective mosaic of the single crystal is about 0.8$^\circ$ with the spectrometer configurations described below. The inelastic neutron scattering experiments were carried out on the thermal triple-axis neutron spectrometer HB-1 and the cold triple-axis neutron spectrometer CTAX, installed at HFIR at ORNL. Neutrons with a final energy of 14.7~meV and 3.5~meV were used, together with a horizontal collimator sequence of $48'$--$80'$--S--$80'$--$120'$ and guide--open--S--$80'$--open on HB-1 and CTAX, respectively. Contamination from higher-order beams was effectively eliminated using PG and Be filters on HB-1 and CTAX, respectively. In both experiments, the single crystal was oriented in the $(H\bar{H}0)$-$(00H)$ scattering plane and was mounted in a closed-cycle $^4$He gas refrigerator.

\begin{figure}
\includegraphics[width=8.0cm]{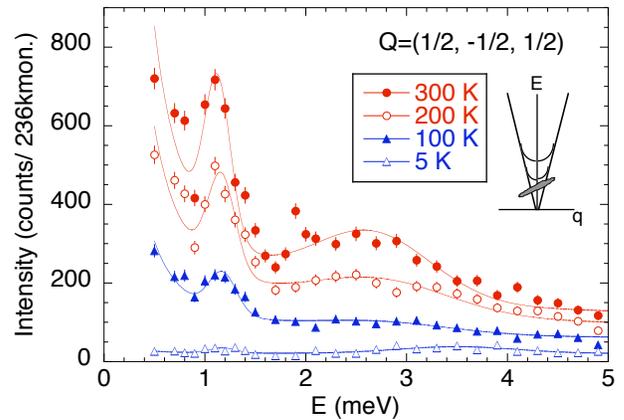}
\caption{(Color online) Low-energy magnetic excitations at the magnetic zone center (1/2, $-$1/2, 1/2) measured at $T$=5, 100, 200, and 300 K on CTAX using cold neutrons. The solid curves are guides to the eye.
The inset is a schematic figure to show the relation between the dispersions and the instrumental resolution (shaded ellipsoid).}
\label{zone_center}
\end{figure}
Figure \ref{spectra} shows the typical inelastic neutron spectra along $[h,\bar{h},h]$, $[h,\bar{h},\bar{h}]$, $[h,\bar{h},0]$, and $[0,0,h]$ in BiFeO$_3$ measured on HB-1. Since the dispersion is steep and the instrument resolution in $Q$ is not sufficient to resolve the dispersions arising from the incommensurate positions, the observed spectra are considered to be almost the same as those expected for the commensurate G-type structure, as shown in the inset of Fig. \ref{zone_center}, and it is difficult to resolve the two peaks at +$q$ and $-q$ below $E$$\sim$20 meV in constant-energy scans. In order to complete the spin-wave dispersion relations, we mostly used constant-energy scans between $\sim$20 and $\sim$60 meV and constant-$Q$ scans between $\sim$10 and $\sim$20 meV and around the magnetic zone boundary. Below 10 meV we used the constant-energy scans measured on CTAX. In order to determine the peak positions of the spin-wave excitations in the energy-$Q$ space, the CTAX data were fitted using the Gaussian function without convoluting with the resolution function. The HB-1 data were fitted using the Lorentzian function $A$/[$(E-E_0)^2$+$\Gamma^2$] with $\Gamma$=1.5 meV, where $A$ and $E_0$ are constant and peak position in energy, respectively, convoluted with the instrumental resolution function. As shown in Fig. \ref{spectra}, the model function reproduces the observed spectra reasonably well. Therefore, the broad peak widths in the constant-$Q$ scans are not intrinsic but primarily originate from the steep dispersion around the zone center ($<$$\sim$30 meV) and from the insufficient instrumental energy resolution at higher transfer energies ($>$$\sim$30 meV).

\begin{figure}
\includegraphics[width=8.5cm]{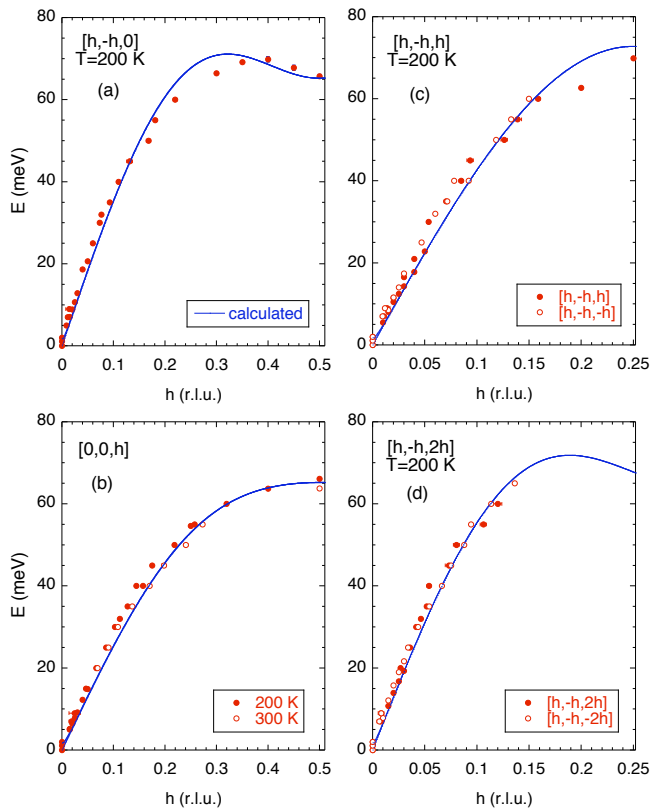}
\caption{(Color online) Magnetic dispersion relations along $[h,\bar{h},0]$ (a), $[0,0,h]$ (b), $[h,\bar{h},h]$ (c), and $[h,\bar{h},2h]$ (d) in BiFeO$_3$. The solid curves are spin-wave dispersion relations calculated with $J_1$=6.48 meV and $J_2$=0.29 meV.
}
\label{dispersions}
\end{figure}

The experiments were carried out primarily at $T$=200 K. Temperature dependence of the dispersion along $[$$0,0,h$$]$ was also measured below 300 K. As shown in Figs. \ref{spectra}(e), \ref{spectra}(f) and \ref{dispersions}(b), the dispersion becomes slightly softened at 300 K with a change of $\sim$2 meV at the zone boundary. The change of dispersion is almost negligible below 200 K (not shown).

Figure \ref{zone_center} shows the temperature dependence of the low energy magnetic excitations at the magnetic zone center (1/2, $-$1/2, 1/2) measured on CTAX.
Because there are several magnetic domains which have different propagation vectors, there are 12 incommensurate magnetic Bragg positions around (1/2, $-$1/2, 1/2), $i. e.$ (1/2$\pm$$\delta$, $-$1/2$\mp$$\delta$, 1/2), (1/2$\pm$$\delta$, $-$1/2, 1/2$\mp$$\delta$), (1/2, $-$1/2$\pm$$\delta$, 1/2$\mp$$\delta$), (1/2$\pm$$\delta$, $-$1/2$\pm$$\delta$, 1/2), (1/2$\pm$$\delta$, $-$1/2, 1/2$\pm$$\delta$), and (1/2, $-$1/2$\pm$$\delta$, 1/2$\pm$$\delta$). Although only the excitations arising from (1/2$\pm$$\delta$, $-$1/2$\mp$$\delta$, 1/2) are observed at higher energies, some of the excitations are superposed around the zone center due to the instrument resolution in $Q$. Even so, gap energies can be determined because they are the same for all domains.
At 300 K, we found a sharp peak at $\sim$1.1 meV and a broad peak at $\sim$2.5 meV.
Those excitation peaks are considered to be magnetic in origin, because they are dispersive and continuously connected to the spin-wave excitations described above. Furthermore, (1/2, $-$1/2, 1/2) corresponds to a zone boundary of the chemical reciprocal lattice unit so that low-energy phonons are not expected to be observed at this position.
Those peaks are broader than the instrumental energy resolution $\sim$0.2 meV around $E$=1$\sim$2 meV. Since the dispersion curve is steep, a slight tail is observed above the gap energy, as shown in the inset of Fig. \ref{zone_center}.
The gap energy of the lower excitation mode is estimated to be $\sim$1.1 meV, whereas that of the higher excitation mode is estimated to be $\sim$2.5 meV, which is difficult to determine more accurately.
As will be described below, these gaps probably originate from the DM interaction and the single-ion anisotropy that give rise to an easy-plane anisotropy in the plane defined by $[1,1,1]$ and the spiral direction $[1,-1,0]$ with a finite in-plane anisotropy along $[1,1,1]$. The scattering intensity below 0.8 meV is considered to originate from the magnetic excitations from the almost gapless mode.
With decreasing temperature, the scattering intensity decreases, following the Bose factor. The lower gap energy just slightly increases, which is consistent with the result that the incommensurability does not change considerably below 300 K. \cite{ramazanoglu11,JHA10,sosnowska11} It is noted that the magnetic anisotropy is small, as expected from the absence of the orbital degree of freedom in the Fe$^{3+}$ ions (3$d^5$, $S$=5/2).

\begin{figure}
\includegraphics[width=8.5cm]{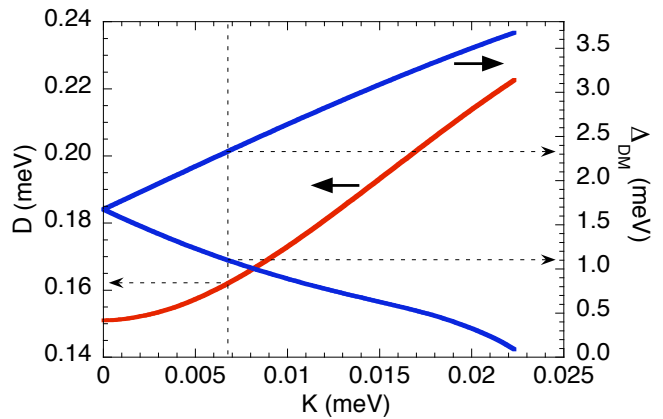}
\caption{(Color online) The relation between $D$ and $K$ that reproduce the cycloidal magnetic structure with a long period of $\sim$640 \AA. The gap energy at the zone center due to the DM interaction $\Delta\rm_{DM}$ is also plotted as a function of $K$.
}
\label{D_K}
\end{figure}
Figure \ref{dispersions} shows the full magnetic dispersion relations of the spin-wave excitations along $[h,\bar{h},0]$, $[0,0,h]$, $[h,\bar{h},h]$, and $[h,\bar{h},2h]$. The maximum energy of the spin-wave excitation is $\sim$70 meV. In order to analyze the spin-wave dispersion relations observed, we assumed the following effective spin Hamiltonian. \cite{jeong12,talbayev11,sosnowska95}

\begin{eqnarray*}
H &=& J_1\sum_{n.n.}\mbox{\boldmath $S$}_i\cdot\mbox{\boldmath $S$}_j+ J_2\sum_{n.n.n.}\mbox{\boldmath $S$}_i\cdot\mbox{\boldmath $S$}_j\\
&-&\mbox{\boldmath $D$}\cdot\sum_{[1,-1,0]}(\mbox{\boldmath $S$}_{i}\times\mbox{\boldmath $S$}_j)-K\sum_i (S_i^{111})^2
\label{Ham}
\end{eqnarray*}
The first and the second terms represent exchange interactions between nearest-neighbor and next-nearest-neighbor spins, respectively. The third and the fourth terms originate from the DM interaction and the single-ion anisotropy, respectively. A moment size of 4.1$\mu_B$ at 200 K was used in the calculations, since the saturated moment was reported to be 4.34$\mu_B$ and it is reduced by $\sim$5\%\ at 200 K.\cite{fischer80}
Calculations for the spin dynamics were performed by including both the DM interaction $D$ and anisotropy $K$ in the Heisenberg Hamitlonian. For a fixed period of the spin helix, $K$ depends on $D$. Since $\delta$ = 0.0045 is approximately equal to 1/222, a unit cell of length 222 was used to evaluate the anharmonic contributions to the spin helix, which was expanded in odd harmonics of a fundamental wavevector $Q$. \cite{fishman10} The spin excitations were evaluated by performing a 1/$S$ expansion in the rotated frame of reference for each spin in the unit cell. The equations-of-motion for the spin operators were solved by diagonalizing a 444-dimensional matrix.  This technique is described in more detail in Ref.\cite{haraldsen09}.  Zone folding then generated the two gap frequencies, which depend on the anisotropy $K$.

We first consider the overall dispersion relations, which are mostly caused by the first and the second terms. The solid curves are spin-wave dispersion relations calculated with $J_1$=6.48 meV and $J_2$=0.29 meV. These values, which describe the observed dispersions reasonably well, are consistent with $J_1$=4.38 meV and $J_2$=0.15 meV from Ref. \cite{jeong12}, where a moment of $2\mu_B\sqrt{S(S+1)}=5.8 \mu_B$ or 1.44 times the $T=200$ K moment was used.
Our results predict the band maximum at 72.8 meV, which is also consistent with 72.5 meV in Ref. \cite{jeong12}. Magnon density of states measured with polycrystalline sample is supposed to show a peak around the band maximum. \cite{mcqueeney08} The peak position was experimentally determined at 68.2 meV at 20 K \cite{sosnowska09} and at 65 meV at 300 K \cite{delaire12}. These are slightly lower than the predicted values probably because the instrumental resolution broadens and lowers the asymmetric peak from the van Hove singularity.

Magnetic anisotropies should be included to explain the excitation gaps around the magnetic zone center, although they do not affect the excitations above $\sim$5 meV. The DM vector is assumed to point along the $[1,1,-2]$ direction and the summation was done for spin pairs along the $[1,-1,0]$ direction. It was reported that there is a finite easy-axis anisotropy along the $[1,1,1]$ direction due to the single-ion anisotropy. \cite{ramazanoglu11,zalesskii02} These terms give rise to excitation gaps around the magnetic zone center shown in Fig. \ref{zone_center}. $D$ and $K$ cannot be chosen independently to reproduce the cycloidal magnetic structure with a long period of $\sim$640 \AA\ previously observed. \cite{sosnowska82,lebeugle07}
Furthermore, the single-ion anisotropy lifts the degeneracy of the excited state.
The relation between $D$ and $K$ is plotted in Fig. \ref{D_K}. In the region of $K$$>$0.0222, the collinear G-type magnetic structure becomes stable. Since the lower excitation gap, which is sharp and well-defined, is observed at 1.10$\pm$0.05 meV, $K$ is estimated to be 0.0068$\pm$0.0007 meV. Then, $D$ is estimated to be 0.1623$\pm$0.0022 meV. Consequently, the higher excitation is predicted to be at 2.33 meV, which is comparable to the experimental result, although the exact peak position is difficult to locate experimentally.
The gap energy due to the easy-axis anisotropy $K$ is predicted to be less than 0.1 meV. Therefore, it is difficult to observe the anisotropy gap experimentally.
The magnetic excitation observed below 0.8 meV corresponds to this spin-wave mode.
The lower gap energy is considered to change linearly with the Fe moment, which decreases by $\sim$5\%\ from 100 to 300 K. \cite{fischer80,ramazanoglu11} This is consistent with the behavior that the lower gap energy slightly decreases by $\sim$3\%\ from 100 to 300 K, as shown in Fig. \ref{zone_center}.

The $K$ term gives rise to an anharmonicity of the cycloidal structure, which was previously suggested by NMR and neutron diffraction studies,~\cite{zalesskii02,ramazanoglu11,sosnowska11} Higher harmonics of the incommensurate magnetic peaks are induced due to the anharmonicity. From $K$=0.0068 meV, the ratio of the intensities of the first to the third harmonics is estimated to be $I_1/I_3$$\sim$120,
which is between $\sim$500 at $T$=5 K \cite{ramazanoglu11} observed in the neutron diffraction measurement and $\sim$25 at $T$=5 K estimated from the NMR measurement.~\cite{zalesskii02} Further studies are required to understand the discrepancy between these measurements.

It is interesting that the zone-center gaps around 1 and 2.5 meV are consistent with the magnon modes observed in Raman scattering and terahertz spectroscopy measurements,~\cite{singh08,cazayous08,rovillain08,talbayev11,sausa08} in which the modes are considered to be electromagnons.
This indicates that the zone-center spin-wave excitations were observed previously, although the electromagnon modes are observable at $q$=0 in principle.
Our results give important information to understand the coupling in those measurements.

The direct observation of the excitation gaps enables us to determine an accurate value for the DM interaction. The presence of the DM interaction suggests that the additional ferroelectric polarization below $T\rm_N$ is caused by the magneto-electric effect predicted by the spin current mechanism. \cite{katsura05}
Therefore, the cycloidal spin structure in BiFeO$_3$ is not caused by geometrical magnetic frustration as in many other multiferroic materials but rather by the relatively large DM interaction.  Hence, perturbations that increases the DM term should enhance the spontaneous electric polarization.

In summary, from the direct observation of the excitation gaps, we determined all the magnetic coupling parameters, including the DM interaction and the single-ion anisotropy, which are consistent with the observed anharmonicity of the cycloidal magnetic structure in multiferroic BiFeO$_3$. Based on the DM interaction, we conclude that
the additional ferroelectric polarization below $T\rm_N$ is caused by the sizable magneto-electric effect.

We would like to thank Prof. N. Furukawa for stimulating discussions. The work at ORNL was sponsored by the Scientific User Facilities Division (MM) and Materials Sciences and Engineering Division (RF), Office of Basic Energy Sciences, U. S. Department of Energy. The work at AIST was partly sponsored by the Funding Program for World-Leading Innovative R\&D on Science and Technology (FIRST Program), Japan.

\end{document}